\documentclass[twocolumn,epjc3]{svjour3}

\usepackage[T1]{fontenc}
\usepackage[utf8]{inputenc}

\usepackage{graphicx,amssymb,amsmath}
\usepackage{hyperref}

\def \d {{\rm d}}
\def \e {e}

\def \boldl {\mbox{\boldmath$l$}}

\begin{document}

\title{Nonsymmetric Dynamical Thin-Shell Wormhole in Robinson--Trautman Class}

\author{O. Sv\'{\i}tek\thanksref{e1,addr1}
        \and
        T. Tahamtan\thanksref{e2,addr1,addr2}}
\thankstext{e1}{e-mail: ota@matfyz.cz}
\thankstext{e2}{e-mail: tahamtan@utf.mff.cuni.cz}
\institute{Institute of Theoretical Physics, Faculty of Mathematics and Physics, Charles University, V~Hole\v{s}ovi\v{c}k\'ach 2, 180~00 Prague~8, Czech Republic\label{addr1}
\and
 Astronomical Institute, Czech Academy of Sciences, Bo\v{c}n\'{\i} II 1401, Prague, CZ-141 31, Czech Republic\label{addr2}}

\date{\today}

\maketitle

\begin{abstract}
\sloppy
The thin-shell wormhole created using the Darmois--Israel formalism applied to Robinson--Trautman family of spacetimes is presented. The stress energy tensor created on the throat is interpreted in terms of two dust streams and it is shown that asymptotically this wormhole settles to the Schwarzschild wormhole with a throat located at the horizon position. This behavior shows a nonlinear stability (within the Robinson--Trautman class) of this spherically symmetric wormhole. The gravitational radiation emitted by the Robinson--Trautman wormhole during the transition to spherical symmetry is indistinguishable from that of the corresponding black hole Robinson--Trautman spacetime. Subsequently, we show that the higher-dimensional generalization of Robinson--Trautman geometry offers a possibility of constructing wormholes without the need to violate the energy conditions for matter induced on the throat.

\fussy


\end{abstract}

\section{Introduction}

Robinson--Trautman spacetimes, distinguished by the presence of expanding, nontwisting and nonshearing null geodesic congruence \cite{RobinsonTrautman:1960,RobinsonTrautman:1962,Stephanietal:book,GriffithsPodolsky:book}, found numerous uses in studies on nonspherical generalizations of black holes (albeit nonrotating ones only). This class of geometries contains many important solutions as a special case, e.g. Schwarzschild and Vaidya solutions or C-metric. This family is also in general without any symmetries and provides dynamical spacetimes with exact gravitational waves emanating from a compact region. As such it provides grounds for studies of exact solutions beyond spherical or axial symmetry --- testing the robustness of properties derived in the symmetric cases and their stability with respect to nonlinear perturbations within this class.

Wormholes represent one of the more peculiar predictions of general relativity when the energy conditions are relaxed or other more exotic options invoked. They have a relatively long history \cite{Misner and Wheeler,Wheeler} and provide an attractive tool for understanding  the general relativity \cite{Morris and Thorne}. Soon after initial considerations other simple models appeared --- e.g. polyhedral wormholes \cite{Visser:1989} (where the unavoidable energy condition violation is concentrated on the edges) or Reissner-Nordstr\"{o}m wormhole \cite{Harris}. The energy conditions violations for wormholes were investigated \cite{WANG and LETELIER} based on similar studies regarding curvature singularities \cite{energy conditions} and it was realized that although one may shift the energy violation region in space and/or time and there may exist geodesics passing through the wormhole completely avoiding this problematic region the energy condition violation is unavoidable \cite{Hochberg and Visser:flare out}. However, recently a possibility to form a wormhole using only a NUT charge and a nonlinear sigma model matter on an anti--de Sitter background was noted \cite{Canfora}.

There are generally two categories of wormholes, both of them covered in the book dedicated to the subject \cite{book}. In the first one, the asymptotic regions are connected smoothly together via a "bridge" and the corresponding metric is regular everywhere. The matter content is smoothly distributed (usually in the neighborhood of the throat) and on average does not satisfy energy conditions. Here, the exact position of a throat has to be determined in a way similar to quasilocal horizons --- as an anti-trapped surface \cite{Hochberg and Visser:1998} or other closely related concepts \cite{Hayward,Tomikawa}. In the second case, the wormhole is created by cutting and gluing existing regular spacetimes. These so called thin-shell wormholes have a well defined throat due to the process of creation. On such a throat a distributional source is created which does not satisfy energy conditions in order to provide transition from the collapsing region to the expanding one. There are studies of thin shell wormholes in spherical symmetry --- static \cite{Poisson and Visser:stability} or dynamical \cite{general SSD thin shell} (see as well numerous references therein), cylindrical symmetry \cite{cylindrical} or in higher dimensions \cite{d-dimension}. Both concepts of wormholes were used widely in various scenarios, mainly looking for a model with the least violation of energy conditions or studying the stability of wormholes.

The stability of thin-shell wormholes was mainly studied using linearization combined with the effective potential method \cite{Poisson and Visser:stability,stability:2002,stability:2011} and its generalizations, e.g. \cite{stability:2008}. Majority of these studies considered highly symmetric (spherical) situations (and the perturbations were usually limited to purely radial as well). Specifically, there are many linear stability results for a thin-shell wormhole in the Schwarzschild geometry \cite{Varela} (and the references therein) but the present paper offers a first results based on a class of exact solutions without any symmetries thus offering a full nonlinear treatment of the problem. The selected Robinson--Trautman class of spacetimes contains reasonably general (albeit rotationless) deformations of Schwarzschild geometry which makes it an ideal tool for stability analysis of spherically symmetric wormholes without the need to select among several methods for investigating the linear stability. Moreover, this class of spacetimes contains gravitational waves as well.

The thin-shell wormhole construction is based on the Darmois--Israel formalism \cite{Israel} which is widely used in the context of relativistic sources and became part of standard computational tools \cite{Musgrave}. In the following, we will apply it to the Robinson--Trautman family of spacetimes. The main analytical results concerning this family are related to the dynamical evolution towards final spherically symmetric Schwarzschild geometry which is attained exponentially fast (in retarded time) in the most interesting type II subfamily in vacuum --- result obtained by Chru\'{s}ciel and Singleton \cite{Chru1,Chru2,ChruSin}. This result is based on the parabolic type equation (Robinson--Trautman equation --- the only nontrivial Einstein equation for this class in vacuum) governing the evolution from initial data which gives a well defined solution to the future --- analogously to the heat equation (this means that extending the solution to infinite negative retarded time is generally impossible). These results were later generalized to nonvanishing cosmological constant \cite{podbic95,podbic97} --- solutions settle down to a Schwarzschild--(anti-)de Sitter --- and to pure radiation source \cite{BicakPerjes,PodSvi:2005} --- solutions approach the spherically symmetric \mbox{Vaidya} metric. In the standard setting of Robinson--Trautman solution (retarded time version) the location of a past horizon together with its general existence and uniqueness for the vacuum Robinson--Trautman solutions has been studied by Tod \cite{tod}. Later, Chow and Lun \cite{chow-lun} analyzed other properties of this horizon and made numerical study of both the horizon equation and the Robinson--Trautman equation. These results were later extended to nonvanishing cosmological constant \cite{PodSvi:2009}. The existence of the horizon together with its nature was studied in the higher-dimensional generalization of Robinson--Trautman spacetimes  \cite{Svitek2011}. The deformed horizon in Robinson-Trautman spacetime and its associated asymptotic momentum caused by directional gravitational radiation emission was also used in the analytic explanation of an "antikick" appearing in numerical models of asymmetric binary black hole mergers \cite{rezzolla}. Apart from the Maxwellian electrodynamics one can couple the Robinson--Trautman geometry to several models of nonlinear electrodynamics \cite{TS_nonlinear}.

\sloppy
Recently, the metric form of the Robinson--Trautman spacetimes was amended in such a way that it admits more general types of matter. Namely, there is a solution containing free massless scalar field \cite{TS-1scal} which was later shown to include many important spherically symmetric scalar field spacetimes \cite{TS-2scal}. The ghost field (scalar field with negative kinetic term in the Lagrangian) version of the spacetime was investigated as well and it was shown that it contains a wormhole throat existing for a finite time \cite{TS-2scal}. This is an example of a bridge-type wormhole with regular geometry and matter.

\fussy


The higher-dimensional generalization of Robinson--Trautman spacetimes \cite{podolsky-ortaggio} presents an opportunity to study compact wormhole throats with less geometrical restrictions (compared to automatic $S^{2}$ topology in four dimensions) similarly to the case of black holes in higher dimensions. Thin-shell wormholes in higher dimensions were already employed in various scenarios \cite{Rahaman,Thibeault}. In section \ref{section6} we will show that the possibility to have compact Einstein manifolds with nonpositive curvature allows to construct wormholes without the need for matter of negative energy density using the higher-dimensional Robinson--Trautman solution.

The below presented investigation is as well quite a timely one since according to recent paper \cite{PRL-QNM} the early ringdown phase emission of gravitational waves detected recently \cite{LIGO} does not hold enough data to reliably distinguish between true black hole signal and that of some other potential compact source even-though later the quasinormal modes differ drastically. Since our model describes fully nonlinear dynamical transition of a compact object (a wormhole --- case considered in \cite{PRL-QNM} as well) to the final stationary phase it represents an example of such a process treatable nonperturbatively.

\section{Vacuum Robinson--Trautman metric and field equations}
\label{RTmetricsec}
\sloppy
The general form of a vacuum Robinson--Trautman spacetime (with cosmological constant) has the following form implementing the twistfree and shearfree conditions \cite{RobinsonTrautman:1960,RobinsonTrautman:1962,Stephanietal:book,GriffithsPodolsky:book}
\begin{equation}\label{RTmetric}
\d s^2 = -2H\,\d u^2-\,2\,\d u\,\d r + \frac{r^2}{{P}^2}\,(\d y^{2} + \d x^{2}),
\end{equation}
where ${2H = \Delta(\,\ln {P}) -2r(\,\ln {P})_{,u} -{2m/r} -(\Lambda/3) r^2}$,
\begin{equation}\label{Laplace}
\Delta\equiv {P}^2(\partial_{xx}+\partial_{yy}),
\end{equation}
and $\Lambda$ is the cosmological constant. The geometry is specified by two functions, ${\,{P}(u,x,y)\,}$ and ${\,m(u)\,}$, satisfying the nonlinear Robinson--Trautman equation (the only nontrivial Einstein equation remaining)
\begin{equation}
\Delta\Delta(\,\ln {P})+12\,m(\,\ln {P})_{,u}-4\,m_{,u}=0\,.
\label{RTequationgen}
\end{equation}
The function $m(u)$ might be set to a constant by suitable coordinate transformation for vacuum solution which makes its interpretation as a quantity related to the mass problematic. However, it appears in formula for the Bondi mass of the spacetime together with the function $P$.

The spacetime is defined as possessing a geodesic, shearfree, twistfree and expanding null congruence. In the case of the metric (\ref{RTmetric}) this congruence is generated by ${\boldl=\partial_r}$. The coordinate $r$ is clearly an affine parameter along this congruence, while $u$~is a retarded time coordinate, spatial coordinates $x,y$ span transversal 2-spaces with their Gaussian curvature (for ${r=1}$) being given by
\begin{equation}
{K}(u,x,y)\equiv\Delta(\,\ln {P})\,.
\label{RTGausscurvature}
\end{equation}
We assume that the transversal 2-spaces are compact (leading necessarily to the spherical topology) which is the standard view of this family of spacetimes. For general fixed values of $r$ and $u$, the Gaussian curvature is ${{K}/r^2}$ so that, as ${r\to\infty}$, the transversal 2-spaces become effectively locally flat.

\section{Wormhole setup}
In this section we consider gluing together two identical copies of the asymptotic part of Robinson--Trautman spacetime (mirror-like setup) along hypersurface $r=f(u)$ whose spatial sections are necessarily compact. However, since the geometry of these sections together with their evolution in retarded time is specified by metric function $P$ they are not simply spheres and evolve nontrivially. This enables us to study quite general wormhole throat while retaining clear geometrical picture and simple description. The wormhole throat should be above the horizon which is unfortunately impossible to localize precisely in general. However, while proving the existence of horizons (generally with cosmological constant) the bounds on the horizon position were presented in \cite{PodSvi:2009}. Here we assume that the cut is above the upper bound on horizon location which is given by the minimum of Gaussian curvature (if $\Lambda=0$, as we will assume in the following) as $2m/K_{min}$ (note that $K_{min}=1$ for spherical symmetry --- Schwarzschild horizon).

While having specified the gluing in a way that the induced metric is automatically identical from borh sides of the throat we have to check what kind of matter got induced on the throat by a potential discontinuity in derivatives. In order to do that, we have to compute extrinsic curvature on the throat (essentially from both sides but here we have a symmetric situation) to derive the induced stress energy tensor according to standard Darmois--Israel formalism \cite{Israel}. 

Denoting the normal to the hypersurface by $n_{a}$ we have the induced metric on the throat (\cite{extrinc curvature1} and \cite{extrinc curvature2})
\begin{equation}\label{h}
h_{ab}=g_{ab}-n_{a}n_{b}
\end{equation}
and the extrinsic curvature
\begin{equation}\label{ext-curvature}
\mathcal{K}_{ab}=h^{c}_{a}h^{d}_{b}{\nabla}_{c}n_{d}
\end{equation}
Using the quantities relevant to our case, namely the metric (\ref{RTmetric}), we have the following form of the normal to the throat $$n_{a}=(-\epsilon f_{,u},\epsilon,0,0).$$
From $n_{a}n^{a}=1$ one obtains the normalization factor $\epsilon=(2H+2f_{,u})^{-1/2}$. From the definition (\ref{h}) one obtains an induced metric on the throat (when expressed in the full spacetime coordinates it is necessarily degenerate)
\begin{align}
h_{ab}\d x^{a}\d x^{b}=&-(2H+\epsilon^2\,f_{,u}^2)\,\d u^2-\,2\,(1-\epsilon^2\,f_{,u}^2)\d u\,\d r\nonumber \\
&+\frac{f^2}{{P}^2}\,(\d y^{2} + \d x^{2})-\epsilon^2\,\d r^2,
\end{align}
Using the definition (\ref{ext-curvature}) one can straightforwardly compute $\mathcal{K}_{ab}$ but since the expression for the whole tensor is quite large we first introduce the following 3-dimensional frame
\begin{equation}\label{frame}
  \partial_{\tau}=\epsilon\, \partial_{u}+\epsilon f_{,u}\partial_{r}\ , \partial_{x}\ , \partial_{y}
\end{equation}

adapted to the throat (removing the trivial orthogonal direction and passing from 4-dimensional latin to 3-dimensional greek indices). The first frame vector is a fourvelocity of static observer on the throat. In this frame on the throat of the wormhole the induced metric takes a simple form

\begin{equation}
h_{\mu\nu}\d x^{\mu}\d x^{\nu}=-\d \tau^2 + \frac{f^2}{{P}^2}\,(\d y^{2} + \d x^{2})
\end{equation}
However, generally one cannot integrate the coordinate $\tau$ (proper time of an observer sitting at the fixed position on the throat) so in (\ref{frame}) one should not understand $\partial_{\tau}$ as a coordinate vector but only as a handy notation for a frame vector (and $\d\tau$ for a corresponding covector). This of course means that we cannot transform the $u$-dependence into the $\tau$-dependence in the metric functions even on the throat. If we would retain the $u$ coordinate for the throat description we will obtain the induced metric
\begin{equation}
	-\frac{1}{\epsilon^2}\d u^2 + \frac{f^2}{{P}^2}\,(\d y^{2} + \d x^{2})\ .
\end{equation}
In the following we will adopt a hybrid approach by retaining the frame (\ref{frame}) which has cleaner physical interpretation while using the coordinate $u$ for expressing the individual frame components of tensorial objects.

Using this strategy we compute frame components of the extrinsic curvature
\begin{equation}
	\mathcal{K}_{\mu\nu}=\mathcal{K}_{ab}\,\partial^{a}_{\mu}\,\partial^{b}_{\nu}
\end{equation}
where $\partial^{a}_{\mu}$ denotes components of the $\mu$-th vector of our new frame expressed in the original vector basis corresponding to the original coordinates of the full spacetime. Thus we obtain the following nonzero components of the extrinsic curvature (\ref{ext-curvature})
\begin{align}\label{ExtCurvature}
\mathcal{K}_{\tau\tau}=&-\left.\frac{1}{2\sqrt{2}}\frac{\left(3H_{,r}f_{,u}+2HH_{,r}+H_{,u}+f_{,uu}\right)}{(H+f_{,u})^\frac{3}{2}}\right\rvert_{r=f(u)} \nonumber\\
\mathcal{K}_{\tau x}=&\left.\mathcal{K}_{x\tau}=-\frac{1}{2}\frac{H_{,x}}{H+f_{,u}}\right\rvert_{r=f(u)} \\
\mathcal{K}_{\tau y}=&\left.\mathcal{K}_{y\tau}=-\frac{1}{2}\frac{H_{,y}}{H+f_{,u}}\right\rvert_{r=f(u)} \nonumber\\
\mathcal{K}_{xx}=&\left.\mathcal{K}_{yy}=
\frac{r}{P^2\sqrt{2}}\frac{\left(f_{,u}+2H+r\frac{P_{,u}}{P}\right)}{\sqrt{H+f_{,u}}}\right\rvert_{r=f(u)} \nonumber
\end{align}
Obviously we need to satisfy 
\begin{equation}\label{square-root}
	H+f_{,u}>0
\end{equation}
because of the square root in the first and last equation of (\ref{ExtCurvature}). Let us determine the sign of $\mathcal{K}_{xx}$ which will be important for the following. Its sign is determined by the sign of the bracket in the nominator (see (\ref{ExtCurvature})). Considering (\ref{square-root}), the explicit form of $H$ and a natural assumption $r>0\Rightarrow f(u)>0$ the sign of $\mathcal{K}_{xx}$ depends on the value of the quantity
\begin{equation}\label{sign-quantity}
	\frac{K}{2}-\frac{m}{f}
\end{equation}
where $K=\Delta(\,\ln {P})$. From already mentioned bounds on horizon position \cite{PodSvi:2009} combined with our decision to prevent the wormhole throat from getting below horizon (the last inequality)
\begin{equation}
	\frac{2m}{K_{max}}\leq r_{horizon} \leq\frac{2m}{K_{min}}\leq f
\end{equation}
we derive the following inequality
\[\frac{K}{2}-\frac{m}{f} \geq \frac{K}{2}-\frac{K_{min}}{2} \]
where the right side is always nonnegative so 
\begin{equation}\label{inequality}
  \frac{K}{2}-\frac{m}{f} \geq 0
\end{equation}
This means that $\mathcal{K}_{xx}$ is positive which will become important while discussing the stress energy tensor later on.

\fussy


\section{Surface stress energy}
Now we will present the resulting stress energy tensor $S_{\mu\nu}$ (greek indices enumerate the adapted frame vectors (\ref{frame}) or covectors, as opposed to full spacetime indices denoted by latin letters) induced on the throat hypersurface according to Darmois--Israel formalism
\begin{equation}\label{induced}
8\pi S_{\mu\, \nu}=tr[\mathcal{K}]h_{\mu\, \nu}-[\mathcal{K}_{\mu\, \nu}]
\end{equation}
which gives the following nonzero components in the frame adapted to the throat
\begin{eqnarray}\label{S-compts}
S_{\tau\tau}&=&-\frac{4P^2}{f^2}\mathcal{K}_{xx} \nonumber\\
S_{\tau x}&=&-2\mathcal{K}_{\tau x}\\
S_{\tau y}&=&-2\mathcal{K}_{\tau y} \nonumber\\
S_{xx}&=&S_{yy}=2\mathcal{K}_{xx}-\frac{2f^2}{P^2}\mathcal{K}_{\tau \tau} \nonumber
\end{eqnarray}
Note that the Darmois--Israel formalism automatically ensures that the stress energy tensor is conserved. The main complication when deriving thin-shell wormhole is connected with interpreting the resulting stress energy tensor in terms of some reasonable matter localised on the wormhole throat. We will try to give the interpretation in terms of two perfect fluid streams with stress energy tensor generally having the following form
\begin{equation}
T_{\mu \nu}=\sum_{i=1}^2 [(\rho_{i}+p_{i})v_{i\mu}v_{i\nu}+p_{i}\,h_{\mu\nu}]
\end{equation}
The explicit components are then (we consider dust for simplicity --- $p_{i}=0$)
\begin{eqnarray}\label{T-compts}
T_{\tau\tau}&=&\rho_{1}v_{1{\tau}}^2+\rho_{2}v_{2{\tau}}^2 \nonumber\\
T_{\tau x}&=&\rho_{1}v_{1{\tau}}v_{1x}+\rho_{2}v_{2{\tau}}v_{2x} \nonumber \\
T_{\tau y}&=&\rho_{1}v_{1{\tau}}v_{1y}+\rho_{2}v_{2{\tau}}v_{2y} \\
T_{xx}&=&\rho_{1}v_{1x}^2+\rho_{2}v_{2x}^2 \nonumber \\
T_{yy}&=&\rho_{1}v_{1y}^2+\rho_{2}v_{2y}^2  \nonumber \\
T_{xy}&=&\rho_{1}v_{1x}v_{1y}+\rho_{2}v_{2x}v_{2y}\nonumber
\end{eqnarray}
and we assume velocity normalization for both streams (there is no summation over $i$)
\begin{equation}\label{UC}
v_{i}^{{\tau}}v_{i{\tau}}+v_{i}^{x}v_{ix}+v_{i}^{y}v_{iy}=-1
\end{equation}
Since we need $T_{xx}=T_{yy}$ and $T_{xy}=0$ according to (\ref{S-compts}) we obtain 
\begin{equation}
v_{1x}v_{2x}=-v_{1y}v_{2y} \,\,\, , \,\,\,\,\, \rho_{2}=\rho_{1}\frac{v_{1y}^2}{v_{2x}^2}
\end{equation}
Looking at the form of (\ref{T-compts}) we immediately realize that components of fluid stress energy tensor satisfy the following relation
\begin{equation}
(T_{\tau x})^2+(T_{\tau y})^2=T_{\tau\tau}T_{xx} 
\end{equation}
which in turn limits our wormhole parameters via the following constraint
\begin{equation}\label{constraint}
(\mathcal{K}_{\tau x})^2+(\mathcal{K}_{\tau y})^2=2\mathcal{K}_{xx}(\mathcal{K}_{\tau\tau}-\frac{P^2}{f^2}\mathcal{K}_{xx})
\end{equation}
which (for a fixed geometry) determines the evolution of the function $f$.

\section{Asymptotic behavior and stability}
\sloppy
As mentioned in the introduction, Robinson--Trautman spacetimes asymptotically approach the corresponding spherically symmetric members of the family. Since our wormhole solution is generated using the Robinson--Trautman geometry it must satisfy (away from the throat) the Einstein equations which reduce to the single equation (\ref{RTequationgen}) where we can assume $m=const.$ without loss of generality. This equation was analysed by Chru\'{s}ciel and Singleton \cite{Chru1,Chru2,ChruSin} to obtain asymptotic form of function $P(u,x,y)$ for smooth initial data on hypersurface $u=const.$ which determines the geometry. In our case with $\Lambda=0$ and without null radiation field the final state corresponds to the Schwarzschild metric. Cutting the spacetime to obtain a wormhole does not change the asymptotic behaviour since the geometry is essentially fixed in the $r$ direction. Thus having established the behavior far from the throat one can extend the solution up to the wormhole throat uniquely provided one stays in the Robinson--Trautman family.

\begin{figure}[h]
\begin{center}
\includegraphics[scale=0.74,bb=0 0 320 201]{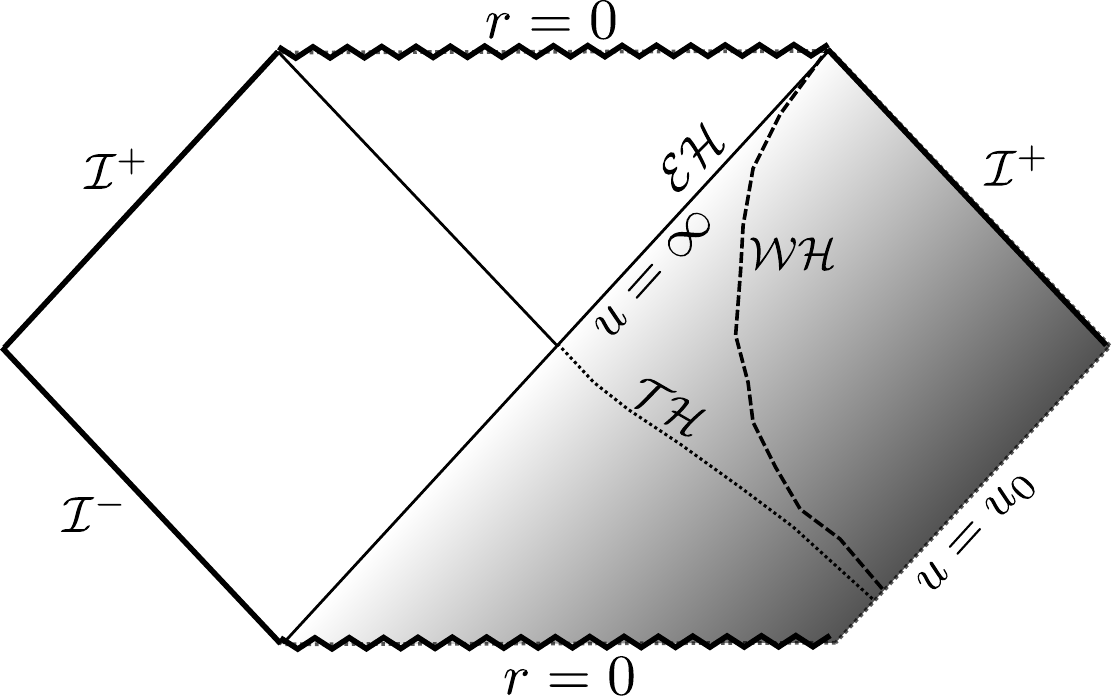}
\end{center}
\caption{\label{figure1}%
Schematic Penrose conformal diagram of Robinson--Trautman spacetime (the shaded portion) which exists for any smooth initial data given on ${u_{0}}$. For ${u\to \infty}$ it approaches the spherically symmetric Schwarzschild solution by emitting gravitational radiation towards future null infinity, and can be extended through the future event horizon~${\mathcal{EH}}$ to the interior Schwarzschild solution and the other asymptotic region (the white portion). The curvature singularities are at ${r=0}$, thick lines represent future and past null infinities ${{\cal I}^+}$ and ${{\cal I}^-}$. The white hole is localized by a trapping (apparent) horizon ${\mathcal{TH}}$ indicated by dotted line. The wormhole throat ${\mathcal{WH}}$ (dashed line) sits above this horizon and asymptotically approaches the future Schwarzschild horizon at $u=\infty$. In the wormhole case only the portion of the diagram to the right from the throat (dashed line) is applicable and is glued to its copy along the dashed line.}
\end{figure}

Now, we will investigate what are the consequences of the asymptotic behavior of Robinson--Trautman solutions for our wormhole. Specifically, the Chru\'sciel's analysis considered the following form of function $P$
\begin{equation}\label{PpP0}
P=p(u,x,y)\, P_{0}\,,
\end{equation}
with $P_{0}=1+\frac{1}{4}(x^2+y^2)$ corresponding to spherical symmetry (with Gaussian curvature $\Delta \ln P_{0}=1$). Using this form in the equation (\ref{RTequationgen}) while assuming $m(u)=const$ (which is always possible to arrange by a suitable coordinate transformation preserving the form of metric) he was able to prove the following asymptotic behavior of $p$ for large values of $u$
\begin{align}
p=& \sum_{i,j\ge0} p_{i,j} u^j \e^{-2iu/m}\nonumber\\
 =&\ 1+p_{1,0}\,\e^{-2u/m}+p_{2,0}\,\e^{-4u/m}+\cdots+p_{14,0}\,\e^{-28u/m} \nonumber\\
&+p_{15,1}\,u\,\e^{-30u/m}+p_{15,0}\,\e^{-30u/m}+\cdots \,,   \label{asymptot}
\end{align}
where $p_{i,j}$ are smooth functions of the spatial coordinates $x,y$ which encode the deviations from spherical symmetry. For large retarded times~$u$, the function $P$ given by \eqref{PpP0} approaches ${P_0}$ exponentially fast and the whole solution thus settles to spherical symmetry (Schwarzschild geometry). The $u=\infty$ hypersurface then corresponds to the future horizon of Schwarzschild black hole and one can attach the inner Schwarzschild solution and the complementary asymptotic region there (see Figure \ref{figure1}).

\fussy

The above described form of $P$ (\ref{PpP0}) leads to the following expression for $H_{,x}$ which determines the asymptotic behavior of $\mathcal{K}_{x\tau}$ (see (\ref{ExtCurvature}))
\begin{equation}
	H_{,x}=\left[\frac{1}{2}\Delta\ln p-r(\ln p)_{,u}\right]_{,x}\ .
\end{equation}
Using the expression (\ref{asymptot}) one can see that even the dominant term of $H_{,x}$ has an exponential falloff. The same applies to $H_{,y}$ of course while the remaining components of extrinsic curvature appearing in (\ref{ExtCurvature}) do not vanish asymptotically. So in order to analyze the asymptotic form of the constraint (\ref{constraint}) we can put $\mathcal{K}_{\tau x}=0$ and $\mathcal{K}_{\tau y}=0$ to obtain
\begin{equation}\label{Asymp-constraint}
	\mathcal{K}_{\tau\tau}-\frac{P^2}{f^2}\mathcal{K}_{xx}=0\ .
\end{equation}
\sloppy
After substituting the asymptotic expansion for $P$ given by (\ref{PpP0}) and (\ref{asymptot}) into expressions for $\mathcal{K}_{\tau\tau}$ and $\mathcal{K}_{xx}$ given in (\ref{ExtCurvature}) and keeping only the terms that are not exponentially suppressed in (\ref{Asymp-constraint}) (effectively using $H\sim 1/2-m/f$ and $P\sim P_{0}$ --- corresponding to Schwarzschild asymptotic limit) one arrives at the following equation
 \begin{equation}\label{f-equation}
f^3\,f_{,uu}+2f^2\,f_{,u}^2+(f-m)(3ff_{,u}+f-2m)=0
 \end{equation}
And since at $u\sim \infty$ we are essentially in the static regime (because of the exponentially fast suppression of the dynamic evolution) we can put $f(u)=f_{0}=const$ as a first rough approximation leading to $f_{0}=2m$. In other words, the wormhole asymptotically settles onto the position of a horizon of the final Schwarzschild geometry although initially it was setup to be in general above it. 
 
 If we use just an approximation where we assume the quartic terms $f^{3}f_{,uu}$ and $f^{2}f_{,u}^2$ in the equation (\ref{f-equation}) to be small compared to remaining lower order terms, we have
 \begin{equation}
 (f-m)(3ff_{,u}+f-2m)=0
 \end{equation}
In this situation we obtain the following solution (we disregard the solution $f=m$ which is definitely below the horizon in the asymptotic region)
\[f(u)=2\,m\,\left\{{\rm LambertW}(z(u))+1\right\}\]
in which
\[z(u)=\frac{1}{2\,m}\exp\left(-\frac{u+C_{1}}{6\,m}-1\right)\]
An expansion for the special function ${\rm LambertW}(z(u))$ (specifically its principal branch according to \cite{Corless}) when $z$ is approaching zero (corresponding to $u\to \infty$) is given by \cite{Corless}
\[z-z^2+\frac{3}{2}z^3-...\]
which shows again that the asymptotic value is $f=2m$ and that going slightly away from $u=\infty$ moves the throat position $f$ above $2m$ (the position of the Schwarzschild horizon). So we have checked that the horizon really approaches the static value from above in this approximation and we can schematically represent the possible throat position in Figure \ref{figure1}.

The above results can be used to establish a nonlinear stability of the final Schwarzschild wormhole with respect to its nonlinear deformations within the Robinson--Trautman class. Namely, if we deform it even far away from spherical symmetry while keeping the throat above the corresponding horizon position it will settle back exponentially fast to a spherically symmetric wormhole by radiating away all the deformations in the form of gravitational radiation (encoded in the Weyl scalar $\Psi_{4}$ (\ref{Weyl}) displayed in the Appendix). The linear stability analysis of a Schwarzschild wormhole using radial perturbations was extended up to $2m$ only recently \cite{Varela}. 

\fussy

\section{Specific values of fluid parameters }
\sloppy
Here we give values of the parameters for the two streams. First for general mutual stream directions, then for one of the streams being static and finally for specific case of perpendicular streams.

\subsection{General streams}
Velocity components are determined in terms of the time component of the second stream. Then the first component of the first stream velocity becomes
\begin{equation}\label{v1t}
v_{1{\tau}}=\sqrt{\frac{\beta^2 v_{2{\tau}}^2-(v_{2{\tau}}^2-1)}{\beta^2 (2\,v_{2{\tau}}^2-1)-(v_{2{\tau}}^2-1)}}
\end{equation}
in which
\[\beta^2=\frac{f^2\mathcal{K}_{\tau\tau}-P^2\,\mathcal{K}_{xx}}{2\,P^2\,\mathcal{K}_{xx}}.\]
One can check from (\ref{constraint}) that $\beta^2$ is indeed positive. Since $v_{1{\tau}}$ and $v_{2{\tau}}$ are bigger than $1$ then, according to (\ref{UC}), we obtain the following restriction on $\beta$ when using (\ref{v1t})
\[\beta^2 < \frac{v_{2{\tau}}^2-1}{2v_{2{\tau}}^2-1}.\]
The spatial components have the following form
\begin{align}
v_{1x}=&\,\frac{f^2}{2P^2\mathcal{K}_{xx}}\frac{1}{\sqrt{\frac{v_{2{\tau}}^2-1}{\beta^2}-(2v_{2{\tau}}^2-1)}}\left[v_{2{\tau}}\,\mathcal{K}_{\tau y}-\alpha^2\,\mathcal{K}_{\tau x}\right]\nonumber\\
v_{2x}=&\,\frac{f^2}{2P^2\mathcal{K}_{xx}}\left[v_{2{\tau}}\,\mathcal{K}_{\tau x}+\alpha^2\,\mathcal{K}_{\tau y}\right] \nonumber \\
v_{1y}=&\,\frac{-f^2}{2P^2\mathcal{K}_{xx}}\frac{1}{\sqrt{\frac{v_{2{\tau}}^2-1}{\beta^2}-(2v_{2{\tau}}^2-1)}}\left[v_{2{\tau}}\,\mathcal{K}_{\tau x}+\alpha^2\,\mathcal{K}_{\tau y}\right]  \nonumber\\
v_{2y}=&\,\frac{f^2}{2P^2\mathcal{K}_{xx}}\left[v_{2{\tau}}\,\mathcal{K}_{\tau y}-\alpha^2\,\mathcal{K}_{\tau x}\right] 
\end{align}
in which
\[\alpha^2=\sqrt{\frac{v_{2{\tau}}^2-1}{\beta^2}-v_{2{\tau}}^2}.\]
The densities can be expressed in the following way
\begin{eqnarray}
\rho_{1}&=&-\frac{4P^2\mathcal{K}_{xx}}{f^2}\left[\frac{(v_{2{\tau}}^2-1)-\beta^2(2\,v_{2{\tau}}^2-1)}{v_{2{\tau}}^2-1}\right]\nonumber\\
\rho_{2}&=&-\frac{4P^2\,\beta^2}{f^2}\frac{\mathcal{K}_{xx}}{v_{2{\tau}}^2-1}
\end{eqnarray}
and their sign is completely determined by the sign of $\mathcal{K}_{xx}$. As discussed after the equation (\ref{inequality}) $\mathcal{K}_{xx}$ is positive and therefore both densities are negative as expected for a viable wormhole.

\fussy

\subsection{Static stream}
Since the time component of the second stream velocity features as a parameter for expressing the rest of the quantities we wish to explore its most prominent value corresponding to the second stream being static with respect to the selected natural frame on the wormhole throat. This means that the second fluid is comoving with the static observer on the throat. In the limit when $v_{2\tau} \longrightarrow 1$ we obtain 
\begin{eqnarray}
\rho_{1}&=&-\frac{4P^2\mathcal{K}_{xx}}{f^2}\left[1-\lim_{v_{2{\tau}}\to 1}{\frac{\beta^2}{v_{2{\tau}}^2-1}}\right]\nonumber\\
\rho_{2}&=&-\frac{4P^2\mathcal{K}_{xx}}{f^2}\left[\lim_{v_{2{\tau}}\to 1}{\frac{\beta^2}{v_{2{\tau}}^2-1}}\right] \nonumber\\
v_{1\tau}&=&\lim_{v_{2{\tau}}\to 1}{v_{1\tau}}=1
\end{eqnarray}
and the rest of the quantities defining streams are zero 
\[\beta=0,\,\,v_{1x}=0,\,\,v_{2x}=0,\,\,v_{1y}=0,\,\,v_{2y}=0.\]
Looking back at the stress energy tensor components (\ref{T-compts}) we see that only $T_{\tau\tau}$ stays nonzero which means that we need to have Robinson--Trautman wormhole with only $S_{\tau\tau}$ nonzero (see (\ref{S-compts})). This means that the function $H$ is independent of $x$ and $y$ which translates (using (\ref{RTequationgen})) into a Gaussian curvature being constant implying spherical symmetry. Thus restricting only one of the streams to being static automatically leads to a spherically symmetric situation.

\subsection{Perpendicular streams}
Now we fix the mutual relation of the streams so that the spatial components of their velocities are perpendicular with respect to the induced metric on the throat. Namely, we select the case when each stream has only one nonzero spatial component of velocity in the frame of the throat. In this case, the stress energy tensor components of these two fluids satisfy
\begin{equation}
	(T_{\tau x})^2/T_{xx} +(T_{\tau y})^2/T_{yy} =T_{\tau\tau}
\end{equation}
and the velocity components are specified completely (with the single spatial component for each stream derived from normalization condition (\ref{UC})) by
\begin{eqnarray}
v_{1{\tau}}&=&\frac{\mathcal{K}_{\tau x}}{\sqrt{\mathcal{K}_{\tau x}^2-\gamma^2}}\nonumber\\
v_{2{\tau}}&=&\frac{\mathcal{K}_{\tau y}}{\sqrt{\mathcal{K}_{\tau y}^2-\gamma^2}} 
\end{eqnarray}
in which
\[\gamma=\frac{f^2\mathcal{K}_{\tau\tau}-P^2\,\mathcal{K}_{xx}}{f\,P}.\]
The densities are unique as well 
\begin{eqnarray}
\rho_{1}&=&-\frac{2\,P\,\left(\mathcal{K}_{\tau x}^2-\gamma^2\right)}{\gamma\,f}\nonumber\\
\rho_{2}&=&-\frac{2\,P\,\left(\mathcal{K}_{\tau y}^2-\gamma^2\right)}{\gamma\,f}
\end{eqnarray}
and using (\ref{constraint}) their sign again depends only on $\mathcal{K}_{xx}$ and is thus negative. This arrangement still covers the general situation.

In the limit of both $\mathcal{K}_{\tau x}$ and $\mathcal{K}_{\tau y}$ going to zero we obtain
\begin{eqnarray}
\rho_{1}&=&\rho_{2}=-\frac{2P^2\mathcal{K}_{xx}}{f^2} \nonumber \\
v_{1{\tau}}&=&v_{2{\tau}}=1
\end{eqnarray}
corresponding to spherically symmetric situation only.

\section{Bulk electromagnetic fields}
There is a simple generalization of the wormhole given above to the case of Robinson--Trautman spacetimes with different electromagnetic fields satisfying Maxwell and several nonlinear electromagnetic field equations given in \cite{TS_nonlinear}. Since the general form of the metric for these spacetimes is the following
\begin{equation}
	\d s^2 = -(2H+Q(u,r))\,\d u^2-\,2\,\d u\,\d r + \frac{r^2}{{P}^2}\,(\d y^{2} + \d x^{2}),
\end{equation}
with $Q(u,r)$ encoding the effect of an electromagnetic field one can use most of the previous calculations directly by replacing the function $H$ by $\tilde{H}=H+Q/2$. The Darmois-Israel formalism automatically provides correct stress energy tensor compatible with the electromagnetic sources created on the throat by the discontinuity in the electromagnetic field.

As shown in the analysis of the horizon existence \cite{TS_nonlinear} the sub and super-solutions associated with the horizon equation provide bounds on the horizon position in a similar way to the vacuum spacetime. Using $Q_{inf}=\inf_{r\in (0,\infty)}\,[Q(u,r)]$ the upper bound now leads to the following restriction on the position of the throat

\begin{equation}\label{EM-bound}
	r_{horizon}\leq\frac{2m}{K_{min}+Q_{inf}}\leq f\ .
\end{equation}
The sign of $K_{xx}$ and subsequently the negativity of the energy density for matter induced on the wormhole throat is now determined by the quantity
\begin{equation}
	\frac{K+Q}{2}-\frac{m}{f}
\end{equation}
generalizing (\ref{sign-quantity}). Using (\ref{EM-bound}) one can easily prove its nonnegativity again leading to matter with nonpositive energy density created on the throat.
The asymptotic behavior for $u\to\infty$ was shown to lead to the corresponding spherically symmetric solutions which by the construction hold for the thin-shell wormhole constructed using these solutions as well.

\section{Higher dimensions}\label{section6}
One of the key steps in proving (\ref{inequality}) and subsequently the negative mass density for the matter on the throat was the observation concerning the bounds on horizon position in four-dimensional Robinson--Trautman spacetime. Those bounds were derived with the horizon viewed as a deformed sphere (with Gaussian curvature $K=1$ for an exact sphere) which is the consequence of both the topology restrictions and the asymptotics of the Robinson--Trautman spacetime. Since in higher dimensions it is possible to have an embeddable compact spatial hypersurface (of codimension 2) with negative curvature this might provide a venue towards wormhole throats supported by matter with positive energy density. The higher-dimensional generalization of Robinson--Trautman family provides a suitable setting for such models as we shall see below.

Robinson--Trautman spacetimes (possibly containing aligned pure radiation and a cosmological constant $\Lambda$) in any dimension were derived recently in \cite{podolsky-ortaggio} using the geometric condition of possesing a nonshearing, nontwisting, but expanding null geodesic congruence (similarly to the definition of the four-dimensional version of the spacetime \cite{RobinsonTrautman:1960,RobinsonTrautman:1962}). The form of the metric valid in $D$ dimensions is similar to (\ref{RTmetric})

\begin{equation}
 \d s^2=\frac{r^2}{P^2}\,\gamma_{ij}\,\d x^i\d x^j-2\,\d u\d r-2H\,\d u^2 
\end{equation}
with 
\begin{eqnarray}\label{H-functionD}
2H&=&\frac{{\cal R}}{(D-2)(D-3)}-2\,r(\ln P)_{,u}-\\
&&\frac{2\Lambda}{(D-2)(D-1)}\,r^2-\frac{\mu(u)}{r^{D-3}}\nonumber
\end{eqnarray} 
The above metric is dependent on a unimodular spatial $(D-2)$-dimensional metric $\gamma_{ij}(x)$, the function $P(x,u)$ and $\mu(u)$ is a ``mass function'' (we assume $\mu>0$). The rescaled metric $h_{ij}=P^{-2}\gamma_{ij}$ must satisfy the field equation ${\cal R}_{ij}=\frac{{\cal R}}{D-2}h_{ij}$.  In $D=4$ this is always satisfied and ${\cal R}$ (the Ricci scalar of the metric $h$) generally depends on $x^i$. However, in $D>4$ this restriction means ${\cal R}={\cal R}(u)$ and the $h_{ij}$ is an Einstein space. Therefore in higher dimensions the dynamical nature of the Robinson--Trautman family is lost (the general algebraic type is just D). On the other hand, with increasing dimensions there is a huge variety of possible spatial metrics $h_{ij}$ (e.g., for ${\cal R} > 0$ and $5 \leq D-2 \leq 9$ an infinite number of compact Einstein spaces were classified). The evolution equation (corresponding to the Robinson--Trautman equation (\ref{RTequationgen})) is much simpler in higher dimensions
\begin{equation}\label{RT-eq-Ddim} 
(D-1)\,\mu\,(\ln P )_{,u}-\mu_{,u} =\frac{16\pi n^{2}}{D-2}\ ,
\end{equation}
where function $n$ describes the possible aligned pure radiation. 

For the simplicity of exposition we will restrict ourselves to vacuum case with $n=0$ and $\Lambda=0$. Using the reparametrization freedom of coordinate $u$ explained in \cite{podolsky-ortaggio} we can put $\mu$ to a constant and from (\ref{RT-eq-Ddim}) we conclude that $P$ is independent of $u$ as well and necessarily ${\cal R}=const$. which means that the Einstein spaces given by $h_{ij}$ have the same scalar curvature for all $u$. The condition of being Einstein is restrictive in dimension 3 since it automatically means a constant sectional curvature. It provides certain topological restrictions in dimension 4 as well - e.g. the Thorpe's theorem which restricts the manifold's characteristic \cite{Besse,Berger}. However in dimension 5 and higher it is not even clear if arbitrary manifold admits an Einstein metric. At the same time in these dimensions there are no known restrictions coming from the sign of Ricci scalar and, e.g., for any dimension $4k$ ($k\geq 2$) there exist manifolds carrying Einstein metrics of both signs - theorem of Catanese and LeBrun \cite{Berger}. We are also in need of a compactness which brings some restrictions but their full extent is not known. If one restricts to manifolds with constant negative sectional curvature (which are automatically Einstein) the standard examples are compact hyperbolic manifolds of arithmetic type, i.e. those arising from factorization of semisimple Lie groups by a discrete subgroup with certain property \cite{Berger}.

Now the question is if the negative scalar curvature of the $D-2$-dimensional Einstein space with metric $h_{ij}$ provides possibility to have positive energy supporting the throat. Let us again consider gluing of two copies along the hypersurface given by $r=f(u)$. Using the formula for induced stress energy tensor (\ref{induced}) together with straightforward generalization of quantities describing the embedding ($a,b,c$ are indices in the full spacetime while $i,j,k$ are reserved for transversal $D-2$-space with metrics $h$ or $\gamma$)
\begin{equation}
 n_{a}\d x^{a}=-\epsilon f_{,u}\d u + \epsilon\, \d r\ ,\ \partial_{\tau}= \epsilon\,\partial_{u}+ \epsilon f_{,u}\partial_{r}	
\end{equation}
we obtain the following expression for the component proportional to the density
\begin{equation}
	S_{\tau\tau}=-2\frac{P^2}{r^2}\gamma^{ij}K_{ij}
\end{equation}
Using the definition (\ref{ext-curvature}) of extrinsic curvature which is valid in higher dimensions as well we obtain
\begin{equation}
	K_{ij}=\frac{r}{P^2}\frac{2H+f_{,u}}{\sqrt{2H+2f_{,u}}}\gamma_{ij}
\end{equation}
So finally we get
\begin{equation}
	S_{\tau\tau}=-(D-2)\frac{2H+f_{,u}}{\sqrt{2H+2f_{,u}}}\, \frac{1}{r}
\end{equation}
If we want the overall density to be positive and the square root in the denominator defined we need to satisfy the following inequalities
\begin{equation}
	2H +f_{,u}\ <\ 0\ <\ H+f_{,u}
\end{equation}
which imply
\begin{equation}\label{f-constraint}
	-H\ <\ f_{,u}\ <\ -2H
\end{equation}
Such a condition means that $H$ has to be negative. If the Ricci scalar ${\mathcal R}$ of the transversal $D-2$-space would be positive then it is possible to use bounds on the horizon position derived from sub- and super-solutions in \cite{Svitek2011} to show that above the horizon $H$ is positive so the condition (\ref{f-constraint}) can never be satisfied and therefore again the energy density on the throat is negative as in the 4-dimensional case.

But in higher dimensions we are presented with the possibility to have compact section of the throat hypersurface with negative Ricci scalar as discussed above. Returning to the results of \cite{Svitek2011} it was not possible to find the horizon in such a case using the techniques therein. If one would assume the horizon having an exactly same geometry as the transversal spaces (on each $u=const$ hypersurface) then the horizon equation derived in \cite{Svitek2011} (with the horizon position given by $r=R(u, x^{i})$)
\begin{align}
&{\cal R}-{\textstyle \frac{2(D-3)}{D-1}}\Lambda R^{2}-{\scriptstyle (D-2)(D-3)}\frac{\mu}{R^{D-3}}{\scriptstyle -{2(D-3)}}\Delta(\ln R)-\nonumber\\
&-{\scriptstyle (D-4)(D-3)}\,h(\nabla \ln R,\nabla \ln R)= 0 \label{PT}
\end{align}
can be used to show that there cannot be any solution for ${\mathcal R}<0$ (since restricting the horizon to have the geometry of the transversal spaces means $r=R(u)$). This indicates that the restriction on the sign of $H$ which was previously coming from the demand that the throat is above the horizon position is removed because the spacetime might not have a horizon at all. Anyway, no matter what are the additional effects of negative scalar curvature we can easily see from (\ref{H-functionD}) that for the case under consideration ($\mu=const>0$, $\Lambda=0$ and $P(x^{i})$) we have $H<0$ everywhere in spacetime. We can start with value $f(u_{0})$ on initial hypersurface $u=u_{0}$ above any possible horizon then the condition (\ref{f-constraint}) ensures that the position of the throat is increasing since $f_{,u}>0$. If the wormhole should be of any use the complete hypersurface of its throat must be timelike which we ensured by the normalization of $n_{a}$. Now we can use the constraint (\ref{f-constraint}) to understand the possible values of gradient of the throat hypersurface $N_{a}=\nabla_{a}[r-f(u)]$ obtaining
\begin{equation}
	0\ <\ g^{ab}N_{a}N_{b}\ <\ -2H
\end{equation}
This means that depending on the choice of $f(u)$ (satisfying the bounds (\ref{f-constraint}) of course) we can approach an almost null hypersurface throat but generically it stays timelike.

\subsection{Calabi-Yau wormholes}
So far we neglected the case where ${\mathcal R}=0$. By inspecting \cite{Svitek2011} we realize that the discussion of ${\mathcal R}<0$ case given in the last paragraph still applies here. Namely the condition $H<0$ is met in this case as well and (\ref{f-constraint}) can be satisfied. This presents us with one well-known family of geometries for our $D-2$-dimensional transversal space which coincides with the geometry of the throat. These are so called Calabi-Yau spaces which are usually defined as compact Kähler manifolds (manifolds which can be equipped with an almost-complex structure compatible with a symplectic form) with a vanishing first Chern class (topological invariant related to the vector bundles over the given manifold) and Ricci flat metric (${\mathcal R}_{ab}=0$ $\Rightarrow$ ${\mathcal R}=0$) \cite{Joyce}. In complex dimension two the simply connected ones are known as K3 surfaces. In the case of complex dimension three it is conjectured that there is a finite number of families of Calabi-Yau manifolds. These are especially important for string theory compactification program \cite{Green} however here we are not suggesting that the wormhole throats are in any way related to these issues. There are known examples of Ricci flat manifolds in odd dimension as well - e.g. the G2 manifold whose holonomy group is contained within the G2 group and admits a spin structure \cite{Joyce}.

\section{Conclusion and final remarks}
\sloppy
We have presented a construction of thin-shell wormhole spacetime using cutting and gluing Robinson--Trautman spacetime. Due to this construction such wormhole is dynamical and has no symmetries. Asymptotically the throat settles to the position of a Schwarzschild horizon corresponding to the final state of evolution in the considered Robinson--Trautman geometry. This result is in agreement with one branch of behavior for non-thin-shell wormholes with only spherical perturbations allowed \cite{Gonzalez}. Note that due to the behavior of the Robinson--Trautman family mentioned in the introduction we cannot generally extend the wormhole to negative infinite retarded time. The stress energy tensor induced on the wormhole throat by the gluing is interpreted in terms of two perfect fluids. It is shown that the densities of these fluids are necessarily negative. Apart from two general streams moving on the throat we consider also one of the streams to be static and the case where the streams are mutually perpendicular which fixes all the freedom of the model.

The asymptotic behavior of the constructed wormhole can be used to establish a nonlinear stability of the Schwarzschild wormhole within the Robinson--Trautman class of geometries.

The asymptotic behavior of our wormhole is completely analogous to that of the standard Robinson--Trautman spacetime with singularity and horizon. The corresponding gravitational radiation (see the Weyl scalar $\Psi_{4}$ (\ref{Weyl}) in the Appendix) which is given by the metric function $P$ (note that $K=\Delta \ln P$) is then identical for our wormhole and the corresponding standard Robinson--Trautman metric thus confirming the results of \cite{PRL-QNM} on nonperturbative level.

The results were shown to be easily generalized to the case of Robinson--Trautman solutions with bulk electromagnetic field of a Maxwell type or satisfying field equations of several models of nonlinear electrodynamics. Finally, we have used the higher-dimensional generalization of the Robinson--Trautman solution to create wormholes with induced matter of positive energy density with several throat geometries.

\fussy

\begin{acknowledgements}
We would like to thank M. Visser for enlightening
discussions. This work was supported by the grant GA\v{C}R 17-13525S.
\end{acknowledgements}

\appendix
\section{Weyl scalars}
Here we present the Weyl scalars corresponding to metric (\ref{RTmetric}) using the frame (where $i$ is a complex unit)
\begin{equation}
	\mathbf{{k}}=\partial_{r}\ ,\quad \mathbf{{l}}=\partial_{u}-H\partial_{r}\ ,\quad \mathbf{{m}}=\frac{P}{\sqrt{2}r}(\partial_{x}+i\partial{y})
\end{equation}

The only nonzero components of the Weyl spinor (corresponding to algebraic type II spacetime) are following
\begin{align}\label{Weyl}
  \Psi_{2}=&-\frac{m}{r^3}\ ,\nonumber\\
  \Psi_{3}=&-\frac{\sqrt{2}P}{4r^{2}}(K_{,x}-iK_{,y})\ ,\\
  \Psi_{4}=&\frac{1}{4r^{2}}\left[\left\{P^2({\tilde{K}}_{,x}-i{\tilde{K}}_{,y})\right\}_{,x}\right.\nonumber\\&\quad\quad\left. -\,i\left\{P^2({\tilde{K}}_{,x}-i{\tilde{K}}_{,y})\right\}_{,y}\right] ,\nonumber
\end{align}
where ${\tilde{K}}=K-2r(\ln{P})_{,u}\,.$


\end{document}